\documentstyle[multicol,prb,aps,tighten]{revtex}

\begin{document}

\draft

\title{Local scale invariance and strongly anisotropic equilibrium 
critical systems}
\author{Malte Henkel}
\address{Laboratoire de Physique des Mat\'eriaux$^{*}$, 
Universit\'e Henri Poincar\'e
Nancy I, B.P. 239, \\ F - 54506 Vand{\oe}uvre-l\`es-Nancy Cedex, France}

\date{23 October 1996}
\maketitle

\begin{abstract}
A new set of infinitesimal transformations generalizing scale 
invariance for strongly
anisotropic critical systems is considered. It is shown that such a
generalization is possible if the anisotropy exponent $\theta =2/N$, 
with $N=1,2,3\ldots$. Differential equations for the two-point function
are derived and explicitly solved for all values of $N$. Known special
cases are conformal invariance ($N=2$) and Schr\"odinger invariance ($N=1$).
For $N=4$ and $N=6$, the results contain as special cases the exactly known 
scaling forms obtained for the spin-spin correlation function in the axial 
next nearest neighbor spherical (ANNNS) model at its Lifshitz points of first 
and second order.
\end{abstract}
\pacs{PACS numbers: 64.60.-i, 05.20.-y, 11.25.Hf}

\begin{multicols}{2}
\narrowtext
The notion of scale invariance is central to the present understanding
of critical phenomena. Here we are interested in strongly anisotropic
criticality. There are many physical examples of this, like 
critical dynamics and nonequilibrium dynamics,\cite{McKa95} 
domain growth,\cite{Bray95} magnetic systems with competing
interactions\cite{Selk92} or particle reaction systems 
such as directed percolation. By 
definition, these systems are characterized by the condition
that the critical two-point 
functions $\cal C$ transform under rescaling as 
\begin{equation} \label{rescal}
{\cal C}(b r, b^{\theta} t) = b^{-2x} {\cal C}(r,t)
\end{equation}
where $r,t$ label `space' and `time' coordinates, $x$ is a scaling dimension
and $\theta=\nu_{\|}/\nu_{\perp}$ is the anisotropy exponent 
(in many cases, it is also referred
to as the dynamic exponent $z$). In this letter, we confine ourselves to 
strongly anisotropic {\em equilibrium} systems. 

Eq.~(\ref{rescal}) can be rewritten as
\begin{equation}
{\cal C}(r,t) = t^{-2x/\theta} \Phi\left(\frac{r^{\theta}}{t}\right)
\end{equation}
where $\Phi(u)$ is a scaling function. Some information on the form of
$\Phi(u)$ is readily available. For $r=0$, one expects ${\cal C}(0,t)
\sim t^{-2x/\theta}$ and for $t=0$, one expects ${\cal C}(r,0)\sim r^{-2x}$. 
This implies $\Phi(u) \simeq \Phi_0$ for $u\rightarrow 0$ and $\Phi(u) \simeq
\Phi_{\infty} u^{-2x/\theta}$ for $u\rightarrow\infty$, where $\Phi_{0,\infty}$
are generically non-vanishing constants.

Is it possible to obtain more information about $\Phi(u)$ on a general basis
without going back to explicit model calculations ? 

Indeed, this has been affirmatively answered in two cases. First, for 
{\em isotropic} critical systems, that is for $\theta=1$, the extension
of eq.~(\ref{rescal}) to space-dependent rescaling factors $b=b(\vec{r}\,)$
leads to the requirement of {\em conformal invariance} of the correlation
functions.\cite{Poly70} (We are not going to restrict ourselves to two
dimensions and shall thus sidestep the extremely powerful and elegant work
done in $2D$, as initiated in Ref. \onlinecite{Bela84}.) Then the critical 
two-point correlation function is, up to normalization\cite{Poly70}
\begin{equation} \label{KonfZw}
{\cal C}(\vec{r}\,) = \langle \phi_{1}(\vec{r}_1) \phi_{2}(\vec{r}_2) \rangle =
\delta_{x_1, x_2} \, |\vec{r}_1 - \vec{r}_2|^{-2x_{1}}
\end{equation}
where $x_{1,2}$ are the scaling dimensions of the 
(scalar) fields $\phi_{1,2}$ which are assumed to be quasiprimary in the 
sense of Ref. \onlinecite{Bela84}. 

Second, for $\theta=2$, the extension of eq.~(\ref{rescal}) to 
space-time-dependent scaling $b=b(\vec{r},t)$ leads to the
requirement of {\em Schr\"odinger invariance}.\cite{Nied72,Hage72} Since this
corresponds to the `non-relativistic' limit of the conformal 
group,\cite{Baru73}
local fields $\phi_i$ are characterized by two quantum numbers, the scaling
dimensions $x_i$ and the masses ${\cal M}_i \geq 0$. For scalar quasiprimary
fields, the two-point function is, up to normalization\cite{Henk92,Henk94}
\begin{eqnarray}
\lefteqn{ 
\langle \phi_{1}(\vec{r}_1, t_1) \phi_{2}^{*}(\vec{r}_2, t_2) \rangle = 
\delta_{x_1,x_2} \, ( t_1 - t_2 )^{-x_1} \, \cdot
} \nonumber \\
&\cdot&  \, \delta_{{\cal M}_1, {\cal M}_2 } 
\exp\left( -\frac{{\cal M}_1}{2} 
\frac{ (\vec{r}_1 - \vec{r}_2)^2}{t_1 -t_2} \right) 
\label{SchrZw}
\end{eqnarray}
with $t_1 > t_2$. In comparing eqns.~(\ref{KonfZw},\ref{SchrZw}), we note that
the first line of (\ref{SchrZw}) is similar to the conformal invariance result,
while the terms containing the masses reflect the non-relativistic nature
of the problem for $\theta=2$. For $\theta=1$, eq.~(\ref{KonfZw}) is
completely standard and there are quite a few statistical mechanics models with
$\theta=2$ which reproduce (\ref{SchrZw}), 
see Refs. \onlinecite{Henk94,Henk94a}.

What are common features of conformal and Schr\"odinger transformations which
might serve as a basis for generalizing beyond $\theta=1,2$ ? For notational
simplicity, we shall work from now on in two `space' dimensions or one 
`time' and one `space' dimension, respectively, but the generalization to
any number of dimensions is immediate. Working in (complex) light-cone 
coordinates $z=x+i y, \bar{z}=x-iy$, the conformal transformations are
\begin{equation} \label{Moeb}
z \rightarrow z' = \frac{\alpha z +\beta}{\gamma z +\delta} \qquad ; \qquad 
\alpha \delta - \beta \gamma =1
\end{equation}
and similarly for $\bar{z}$. The infinitesimal generators are
$\ell_n = -z^{n+1}\partial_z$ and satisfy the commutation relations
$[\ell_n , \ell_m ] = (n-m) \ell_{n+m}$. The set $\{ \ell_{\pm 1}, \ell_0 \}$
generates the M\"obius transformations (\ref{Moeb}). 
The space-time transformations of the 
Schr\"odinger group are\cite{Nied72,Hage72}
\begin{equation}
t \rightarrow t' = \frac{\alpha t +\beta}{\gamma t +\delta} \;\; , \;\;
r \rightarrow r' = \frac{r + v t + a }{\gamma t + \delta} \label{Nied}
\end{equation}
(with $\alpha\delta-\beta\gamma=1$) 
which contains the Galilei group as a subgroup. 
As is well known from non-relativistic quantum mechanics, the wave function
$\psi(r,t)$ transforms under a unitary {\em projective} 
representation $\cal U$ 
of the Galilei transformation\cite{Levy67}
\begin{equation}
{\cal U}^{-1} \psi(r,t) {\cal U} = \exp\left[\frac{i m}{2} 
\left( v^2 t + 2 vr\right) \right] \psi(r+vt, t)
\end{equation}
where $m\geq 0$ is the mass of the particle. This gives rise to the
Bargmann superselection rule\cite{Levy67,Hage72} already present 
in (\ref{SchrZw}). If a wave function $\psi$ is
characterized by the mass $m\geq 0$, its complex conjugate $\psi^*$ is 
characterized by $-m$. This correspondence between a field $\phi$ and
$\phi^*$ is to be kept when going over to diffusive behaviour 
$m\rightarrow i {\cal M}$. An analogous statement applies to 
the full Schr\"odinger
group.\cite{Hage72,Perr77}  The infinitesimal generators must therefore
contain mass terms and may be written in the form\cite{Henk94}
\begin{eqnarray}
X_{n} &=& -t^{n+1}\partial_t - \frac{n+1}{2}t^n r\partial_r -\frac{n(n+1)}{4}
{\cal M} t^{n-1} r^2 \nonumber \\
Y_{m} &=& -t^{m+1/2}\partial_r -\left(m+\frac{1}{2}\right) {\cal M} t^{m-1/2}r
\nonumber \\
M_{n} &=& -t^n {\cal M} \label{SchGen}
\end{eqnarray}
and the non-vanishing commutators are
\begin{eqnarray}
[X_n, X_m] = (n-m) X_{n+m} \:\: &,& \:\: 
{} [X_n, Y_m] =\left( \frac{n}{2}-m\right)
Y_{n+m} \nonumber \\
{} [X_n, M_m] = -m M_{n+m} \:\; &,& \;\: [Y_n, Y_m] = (n-m) M_{n+m}
\nonumber 
\end{eqnarray}
The set $\{ X_{\pm 1}, X_{0}, Y_{\pm 1/2}, M_0 \}$ generates the transformations
(\ref{Nied}). 

We now specify the conditions under which we shall attempt to consider
an arbitrary value of the exponent $\theta$. These conditions are formulated
as to remain as close as possible to the known situations of either
conformal or Schr\"odinger invariance. \\ ~ \\
{\bf 1.} Since in both cases, M\"obius transformations play a prominent role,
we shall seek space-time-transformations which in the `time' coordinate 
undergoes a M\"obius transformation
\begin{equation}
t \rightarrow t' = \frac{\alpha t + \beta}{\gamma t + \delta} \qquad ; \qquad
\alpha\delta -\beta\gamma =1
\end{equation}
{\bf 2.} The generator for scale transformations should read
$X_0 = -t\partial_t -\frac{1}{\theta}r\partial_r$. \\
{\bf 3.} Spatial translation invariance is required. \\
{\bf 4.} The generators should contain `mass' terms, 
built in analogy to the mass
terms for $\theta=2$ in (\ref{SchGen}). \\
{\bf 5.} We want to use these transformations to derive differential equations
for the two-point functions. We shall require that when applied to a two-point
functions, the generators will yield a {\em finite} number of independent
conditions. Thus the operators applied to the two-point functions should
provide a realization of a finite-dimensional Lie algebra. \\

We now proceed to list the consequences of the above assumptions. The
generator $X_n$, $n=-1,0,1$ of the M\"obius transformations must contain
the term $X_n = -t^{n+1}\partial_t + \cdots$ and thus satisfy the commutation
relations $[X_n, X_m] =(n-m)X_{n+m}$. In order to keep the `conformal'
structure of the transformations, we must require that these commutation
relations are also satisfied by the final generators $X_n$. Then the explicit
form of $X_0$ implies that up to mass terms, $X_n = -t^{n+1}\partial_t -
\theta^{-1}(n+1)t^n r\partial_r$. Next, we study the action of $X_n$ on the
space translation operator $-\partial_r$. We shall write $\theta=2/N$ and 
define, up to mass terms, 
the operators $Y_m = -t^{N/2+m}\partial_r$ with $m=-N/2+k$, $k=0,1,\ldots$.  
The nonvanishing commutators of $X_n$ and $Y_m$ are
\begin{eqnarray}
[X_n, X_m] &=& (n-m) X_{n+m} \label{Algebra} \\ 
{}[X_n, Y_m] &=& \left( N\frac{n}{2} -m \right) Y_{n+m} \nonumber
\end{eqnarray}
In particular, 
$[X_1, Y_{-N/2+k}] = (N-k) Y_{-N/2+k+1}$.
Thus, the repeated action of $X_1$ on $Y_{-N/2}=-\partial_r$ is creating an
infinite set of generators. This can only be truncated if $N=2/\theta$ is
a positive integer, $N=1,2,\ldots$. Therefore the list of possible
values of $\theta$ is
\begin{equation}
\theta = \frac{2}{N} = 2, 1, \frac{2}{3}, \frac{1}{2}, 
\frac{2}{5}, \frac{1}{3}, \ldots
\end{equation}

A few remarks are in order. The `conformal' properties of the tranformations
sit in the `time' direction. It should thus be the temporal degrees of freedom
which render the system critical. Therefore one should expect that the results
for the two-point function to be derived below should apply independently of
whether or not the `spatial' degrees of freedom by themselves
furnish a critical
system. One might think of interchanging the roles of `space' and `time'
coordinates and thus obtain a set of anisotropy exponents $\theta= \frac{1}{2},
1, \frac{3}{2}, 2,\ldots$. To do this, however, one must impose `conformal'
invariance on the spatial degrees of freedom and this means that the spatial
degrees of freedom alone should describe a system at 
a critical point. While that
would be fine for a study of critical dynamics, many other examples of
strongly anisotropic critical systems are not at a static critical point. 
In $(1+1)D$, however, this distinction should not be very important, since
a one-dimensional subsystem with short-ranged interactions cannot order
by itself. 

Finally, we have to see whether it is possible to include mass terms into the
generators $X_n, Y_m$ without spoiling the commutator relations (\ref{Algebra}).
Indeed, this can be done. The details of this calculation
will be presented elsewhere, here we merely quote the result. 
One solution for the generators $X_1$
and $Y_{-N/2+1}$ (which generate the so-called `special' and `Galilei' 
transformations) is
\begin{eqnarray}
X_{1} &=& -t^{2} \partial_t -N t r\partial_r -\alpha r^2 \partial_t^{N-1} 
\nonumber \\
Y_{-N/2+1} &=& -t\partial_r -\frac{2\alpha}{N} r \partial_t^{N-1}
\end{eqnarray}
where $\alpha$ is a dimensionful, in general non-universal, 
constant which parametrizes the mass term. When applying
these generators to a two-point function 
${\cal C} =\langle \phi_1 \phi_2\rangle$ where
the fields are characterized by two quantum numbers: the scaling dimension
$x_i$ and the `mass' $\alpha_i$, consistency can only be achieved if
\begin{equation} \label{ABed}
\alpha_1 = (-1)^N \alpha_2
\end{equation}
We point out that for systems with $N$ even, the distinction between 
$\phi$ and $\phi^*$ becomes unnessary. In principle, it is even possible
to introduce a {\em universal} mass constant $\alpha$ which is the same for
all fields.  
On the other hand, for $N$ odd, the $\alpha_i$ must be kept as peculiar 
quantum numbers of the fields $\phi_i$. To each field $\phi_i$, characterized
by the numbers $(x_i, \alpha_i)$, there is a conjugate field $\phi_i^*$ 
characterized by $(x_i, -\alpha_i)$. Furthermore, it can be checked using
(\ref{ABed}) that the two-particle operators built from the $X_n, Y_m$ provide
on $\cal C$ a realization of the Lie algebra (\ref{Algebra}). 

Two special cases can be easily recognized. For $N=2$, we recover 
the familiar conformal algebra, with $X_n = \ell_n + \bar{\ell}_n$ and
$Y_n = i(\ell_n - \bar{\ell}_n)$, $n=-1,0,1$, provided that the `mass' 
$\alpha = -c^{-2}$ 
(where $c$ is the speed of light, normally set to $c=1$ when introducing 
light-cone coordinates $z,\bar{z}=t\pm\sqrt{\alpha}\,r$). 
For $N=1$, we recover the
generators (\ref{SchGen}) of the Schr\"odinger algebra, with 
$\alpha_i = \frac{1}{2}{\cal M}_i$.  

We are now ready to calculate the two-point function 
explicitly. If $X_n^{(a)}$
is the generator $X_n$ acting on particle $a$, $a=1,2$ 
(and similarly for the
$Y_m$), the two-particle operators are 
$\widetilde{X}_n = X_n^{(1)} + X_n^{(2)}$. 
We are interested in the two-point function
\begin{equation}
G(r_1, r_2 ; t_1, t_2 ) = 
\langle \phi_1 (r_1, t_1) \phi_2^{*} (r_2, t_2)\rangle
\end{equation}
and the covariance of $G$ is expressed through the conditions
(meaning that the $\phi_i$ are quasiprimary\cite{Bela84})
\begin{eqnarray}
\widetilde{X}_{0} G &=& \frac{x_1+x_2}{\theta} G \;\; , \;\;
\widetilde{X}_{1} G = \left( \frac{x_1}{\theta} t_1 + 
\frac{x_2}{\theta} t_2 \right) G \nonumber \\ 
\widetilde{X}_{-1} G &=& \widetilde{Y}_{m} G = 0
\end{eqnarray}
with $m=-N/2, -N/2+1, \ldots, N/2$. We write $t=t_1-t_2$ and $r=r_1 -r_2$.
In addition, we put $\zeta = (x_1 + x_2)/\theta$. The scaling of the
two-point function can be written as
\begin{equation} \label{GSkal}
G = G(r,t) = \delta_{x_1, x_2} \, \delta_{\alpha_1, \alpha_2} \, r^{-2 x_1} 
\, \Omega\left( \frac{t}{r^{2/N}} \right)
\end{equation}
where $\Omega(v)$ satisfies the differential equation
\begin{equation} \label{OmEq}
\alpha_1 \Omega^{(N-1)}(v) - v^2 \Omega'(v) - \zeta v \Omega(v) = 0
\end{equation}
subject to the boundary conditions $\Omega(0) = \mbox{\rm cste.}$ and
$\Omega(v) \sim v^{-\zeta}$ as $v\rightarrow\infty$. 
The general solution (for $N\geq 2$) of equation (\ref{OmEq}) is
\end{multicols}
\widetext
\noindent\rule{20.5pc}{0.1mm}\rule{0.1mm}{1.5mm}\hfill
\begin{equation} \label{Omega1}
\Omega(v) = \sum_{p=0}^{N-2} b_p v^p {\cal F}_p \qquad ; \qquad 
{\cal F}_p = {_{2}F_{N-1}} \left( \frac{\zeta+p}{N}, 1 ; 1+\frac{p}{N}, 
1+\frac{p-1}{N}, \ldots, \frac{p+2}{N} ; \frac{ v^N }{N^{N-2} \alpha_1} \right) 
\end{equation}
\hfill\rule[-1.5mm]{0.1mm}{1.5mm}\rule{20.5pc}{0.1mm}
\begin{multicols}{2}
\narrowtext
\noindent where ${_{2}F_{N-1}}$ is a generalized hypergeometric 
function and the
$b_p$ are free parameters. In order to check the boundary conditions, we recall
the known\cite{Wrig35} asymptotic behaviour of the ${\cal F}_p$. The
leading behaviour for $v\rightarrow\infty$ for each term is of the order
$\exp( A (N-2) v^{N/(N-2)})$, where the constant $A>0$.  
For $N\geq 3$ the condition
\begin{equation} \label{BBed}
\sum_{p=0}^{N-2} b_p \frac{\Gamma(p+1)}{\Gamma\left(\frac{p+1}{N}\right)
\Gamma\left(\frac{p+\zeta}{N}\right) } \left( \frac{\alpha_1}{N^2}\right)^{p/N}
=0
\end{equation}
is sufficient to cancel the entire exponential contribution. Eliminating
$b_{N-2}$, the final
result becomes $\Omega(v) = \sum_{p=0}^{N-3} b_p \Omega_p (v)$, with
$b_0 \neq 0$. The asymptotic behaviour
\begin{equation}
\Omega_p(v) \simeq \left\{ \begin{array}{ll}
v^p & \; ; \; \mbox{\rm $v\rightarrow 0$} \\
\Omega_{\infty} v^{-\zeta} & \; ; \; \mbox{\rm $v\rightarrow\infty$} 
\end{array} \right.
\end{equation}
is found to be in complete agreement with the requested boundary conditions,
where
\end{multicols}
\widetext
\noindent\rule{20.5pc}{0.1mm}\rule{0.1mm}{1.5mm}\hfill
\begin{equation} \label{Omega2}
\Omega_p(v) = v^p {\cal F}_p - 
\frac{\Gamma(p+1)}{\Gamma(\frac{p+1}{N})\Gamma(\frac{p+\zeta}{N})} 
\frac{\Gamma(\frac{N-1}{N})\Gamma(1+\frac{\zeta-2}{N})}{\Gamma(N-1)}
\left(\frac{\alpha_1}{N^2}\right)^{(p+2-N)/N} 
v^{N-2} \, {\cal F}_{N-2}
\end{equation}
\begin{equation}
\Omega_{\infty} = -\left( \frac{\alpha_1}{N^2}\right)^{(\zeta+p)/N}
\frac{\Gamma(\frac{1-\zeta}{N})}{\Gamma(1-\zeta)}
\frac{\Gamma(p+1)}{\Gamma(\frac{p+1}{N})}
\frac{\pi \sin\left(\frac{\pi}{N}(p+2)\right)}{\Gamma(\frac{p+\zeta}{N})
\sin\left(\frac{\pi}{N}(p+\zeta)\right)
\sin\left(\frac{\pi}{N}(\zeta-2)\right) }
\end{equation}
%
%
\hfill\rule[-1.5mm]{0.1mm}{1.5mm}\rule{20.5pc}{0.1mm}
\begin{multicols}{2}
\narrowtext
Eq.~(\ref{GSkal}) together with eqs.~(\ref{Omega1},\ref{BBed}) or 
(\ref{Omega2}) gives the solution to our question. After normalization, $N-3$
of the parameters $b_p$ are still arbitrary.  

It remains to be seen whether there exist 
examples which do reproduce these
predictions. Here, we shall consider the spin-spin correlator in spin systems
with axial next nearest neighbor interactions.\cite{Horn75,Selk92} 
The spin Hamiltonian is
\begin{equation} \label{ANNN}
{\cal H} = -J {\sum_{(i,j)}}' s_i s_j +\kappa J \sum_{i\|} s_{i\|} s_{i\|+1}
\end{equation}
where $s_i$ is a $O(n)$ vector spin and 
the first term ($J>0$) describes nearest neighbor ferromagnetic 
interactions while the second term ($\kappa>0$) contains next-nearest neighbor 
interactions
along a single axis. By definition,\cite{Horn75} 
the meeting point of the paramagnetic, 
ferromagnetic and incommensurable phases of the model is termed a {\em Lifshitz
point} (of first order) and is known to show strongly anisotropic scaling, 
with correlation
length exponents $\nu_{\|}=\nu_{\ell4}$, $\nu_{\perp}=\nu_{\ell2}$ 
measured parallel and perpendicular
to the axis. The anisotropy exponent $\theta=\nu_{\|}/\nu_{\perp}=1/2$ 
independently\cite{Horn75} of the value of $n$. This corresponds to $N=4$. 
The fact that $\theta=\frac{1}{2}$ stays fixed at its mean-field value
may point toward the existence of a hidden symmetry which prevents
its renormalization.\cite{Card96} 

In the $n\rightarrow\infty$ limit one recovers the spherical 
(or ANNNS\cite{Selk92}) 
model and the spin-spin
correlation function $C(r_{\|}, \vec{r}_{\perp}) = 
\langle s_{r_{\|}, \vec{r}_{\perp}} s_{0,\vec{0}} \rangle$ 
{\em at} the Lifshitz point is exactly known in 
$d$ dimensions. The result is\cite{Frac93}
\begin{equation}
C(r_{\|}, \vec{r}_{\perp}) = 
C_0 \, r_{\perp}^{-(d-d_*)} \, 
\Psi\left( \frac{d-d_*}{2}, \sqrt{ \frac{1}{32 c_2} } 
\frac{ r_{\|}^2 }{ r_{\perp} } \right)
\end{equation}
where $C_0$ and $c_2$ are known (non-universal) constants, $d_*$ is the
lower critical dimension 
and $\Psi(a,x) = \sum_{k=0}^{\infty} \frac{(-x)^k}{k!} 
\frac{\Gamma(k/2+a)}{\Gamma(k/2+3/4)}$. On the other hand, for $N=4$
eq.~(\ref{GSkal}) gives $G(r,t) \sim r^{-\zeta/2}\, \Omega(v)$. As for
the scaling function $\Omega(v)$, we have from (\ref{Omega2}) that
for $N=4$
\begin{equation}
\Omega_0(v) = \frac{\Gamma(3/4)}{\Gamma(\zeta/4)} \, \Psi\left(\frac{\zeta}{4},
\frac{v^2}{2\sqrt{\alpha_1}} \right)
\end{equation}
Thus, with the correspondence $t \leftrightarrow r_{\|}$, $r \leftrightarrow
r_{\perp}$ and $\alpha_1 = 8 c_2$, the order parameter 
scaling function for the ANNNS model
{\em at} the first order Lifshitz point is exactly reproduced for 
the parameter value $b_1 =0$. 

Higher order Lifshitz points\cite{Selk92} can be reached by adding 
further axial interaction
terms in (\ref{ANNN}). Second order Lifshitz points correspond to 
$\theta=\frac{1}{3}$ or $N=6$. We have checked that the exactly known
spin-spin correlation function for the ANNNS model\cite{Frac93} does 
agree with the scaling form (\ref{Omega2}).  

A tempting open question is whether the
scaling function of the spin-spin correlator of the ANNNI model at the
Lifshitz point (in\cite{Selk92} $3D$), 
which still corresponds to $N=4$,\cite{Horn75} 
can be described in the
same framework with a different value of $b_1$. Recently, a new asymmetric 
six-vertex model with a $\theta=\frac{1}{2}$ critical point has been 
described.\cite{Albe96} Further examples might be provided by the
superintegrable chiral $N-$state Potts model, where\cite{Baxt89} 
$\nu_{\tau}=2/N$, $\nu_{x}=1$ at the self-dual point 
or else by a non-hermitian quantum chain
obtained from the asymmetric clock model, where\cite{Vesc86} 
$\nu_{x}=0.95(4)$ and $\nu_{\tau}=0.67(4)$. 
The possibility of applying the above scheme
to the KPZ equation,\cite{McKa95} which in $(1+1)D$ has $\theta=\frac{3}{2}$,
seems worth exploring.\cite{Work}  
Finally, it appears possible to extend the present 
approach to yield the scaling forms for the response functions out of
equilibrium (as already checked\cite{Henk94} in a few cases for Schr\"odinger 
invariance) and to higher $n-$point functions. This will be reported elsewhere.
All in all, further explicit model results will be needed in order to gauge
the merits of this or any other general approach to strongly 
anisotropic scaling.  

In conclusion, we have examined a set of infinitesimal transformations which
for $\theta=2/N$, $N=1,2,3,\ldots$ generalize scale invariance. 
We have seen how to
calculate from these the two-point functions for strongly anisotropic 
equilibrium critical systems. Lifshitz points in the ANNNS (spherical) model
apparently provide model examples which realize these transformations.

\end{multicols}


\begin{thebibliography}{99}

\bibitem[*]{ura} Unit\'e de recherche associ\'ee au CRNS no. 155. 

\bibitem{McKa95} A. McKane, M. Droz, J. Vannimenus and D. Wolf (Eds)
{\it Scale Invariance, Interfaces and Non-equilibrium Dynamics}, 
NATO ASI, Vol. B344, Plenum (New York 1995). 

\bibitem{Bray95} A.J. Bray, Adv. Phys. {\bf 43}, 357 (1994). 

\bibitem{Selk92} W. Selke, in C. Domb and J.L. Lebowitz (Eds) {\it Phase 
Transitions and Critical Phenomena}, Vol. 15, Academic Press (New York 1992). 

\bibitem{Poly70} A.M. Polyakov, Sov. Phys. JETP Lett. {\bf 12}, 381 (1970).
See L. Sch\"afer J.Phys. {\bf A9}, 377 (1975) for a full discussion how 
conformal invariance enters into eq. (\ref{KonfZw}). 

\bibitem{Bela84} A.A. Belavin, A.M. Polyakov and A.B. Zamolodchikov, Nucl.
Phys. {\bf B241}, 333 (1984). 

\bibitem{Nied72} U. Niederer, Helv. Phys. Acta {\bf 45}, 802 (1972).

\bibitem{Hage72} C.R. Hagen, Phys. Rev. {\bf D5}, 377 (1972).

\bibitem{Baru73} A.O. Barut, Helv. Phys. Acta {\bf 46}, 496 (1973). 

\bibitem{Henk92} M. Henkel, Int. J. Mod. Phys. {\bf C3}, 1011 (1992).

\bibitem{Henk94} M. Henkel, J. Stat. Phys. {\bf 75}, 1023 (1994). 

\bibitem{Henk94a} M. Henkel and G.M. Sch\"utz, Int. J. Mod. Phys. {\bf B8}, 
3487 (1994). 

\bibitem{Levy67} J.-M. Levy-Leblond, Comm. Math. Phys. {\bf 4}, 157 (1967); \\
D. Giulini, Ann. of Phys. {\bf 249}, 222 (1996).  

\bibitem{Perr77} M. Perroud, Helv. Phys. Acta {\bf 50}, 233 (1977). 

\bibitem{Wrig35} E.M. Wright, Proc. London Math. Soc. {\bf 46}, 389 (1940); 
J. London Math. Soc. {\bf 27}, 256 (1952) erratum. 

\bibitem{Horn75} R.M. Hornreich, M. Luban and S. Shtrikman, Phys. Rev. Lett. 
{\bf 35}, 1678 (1975); Phys. Lett. {\bf 55A}, 269 (1975). 

\bibitem{Card96} J.L. Cardy, priv. comm. 

\bibitem{Frac93} L. Frachebourg and M. Henkel, Physica {\bf A195}, 577 (1993).

\bibitem{Albe96} G. Albertini, S.R. Dahmen and B. Wehefritz, J. Phys. 
{\bf A29}, L369 (1996) and cond-mat/9606137.

\bibitem{Baxt89} R.J. Baxter, J. Stat. Phys. {\bf 57}, 1 (1989); 
G. Albertini, B.M. McCoy, J.H.H. Perk and S. Tang, Nucl. Phys. {\bf B314}, 741
(1989);
J.L. Cardy, Nucl. Phys. {\bf B389}, 577 (1993); 
H. Au-Yang and J.H.H. Perk, J. Stat. Phys. {\bf 78}, 17 (1995); 
G.v. Gehlen, hep-th/9606001.

\bibitem{Vesc86} T. Vescan, G.v. Gehlen and V. Rittenberg, J. Phys. {\bf A19},
1957 (1986).  

\bibitem{Work} Comparison with the scaling function for $N=3$ 
would imply an interchange of time and
space, as discussed above. This interchange might not be trivial, since the
non-linearity of the KPZ equation does not enter into the spatial correlations 
(J. Krug, priv. comm.). Work along these lines is in progress.
\end{thebibliography}
\end{document}